\documentclass[12pt, anon]{l4dc2024}

\title[Data-Driven Identification of Attack-free Sensors in Networked Control Systems]{Data-Driven Identification of Attack-free Sensors in Networked Control Systems}
\usepackage{times}

\usepackage{amsmath,amssymb,amsfonts}
\usepackage{lipsum}
\usepackage{color}
\usepackage{wrapfig}
\usepackage{multicol}
\usepackage{cuted}
\usepackage{algorithm}
\usepackage{algpseudocode}
\usepackage{array}
\usepackage{mathtools}
\mathtoolsset{showonlyrefs}
\usepackage{caption}
\usepackage{subcaption}
\newtheorem{assumption}{Assumption}
\newtheorem{problem}{Problem}
\newtheorem{prop}{Proposition}

\usepackage{todonotes}

\makeatletter
\newcommand{\ssymbol}[1]{^{\@fnsymbol{#1}}}
\makeatother
\usepackage{arydshln}
\makeatletter
  \renewcommand*\env@matrix[1][*\c@MaxMatrixCols c]{%
    \hskip -\arraycolsep
    \let\@ifnextchar\new@ifnextchar
  \array{#1}}
\makeatother
\author{%
 \Name{Sribalaji C. Anand} \Email{sribalaji.anand@angstrom.uu.se}\\
 \addr Department of Electrical Engineering, Uppsala University, Uppsala, Sweden.
 \AND
 \Name{Michelle S. Chong} \Email{m.s.t.chong@tue.nl}\\
 \addr Department of Mechanical Engineering, Eindhoven University of Technology, The Netherlands.%
  \AND
 \Name{Andr{\'e} M. H. Teixeira} \Email{andre.teixeira@it.uu.se}\\
 \addr  Department of Information Technology, Uppsala University, Uppsala, Sweden.
}

\begin{document}
\maketitle
\begin{abstract}
This paper proposes a data-driven framework to identify the attack-free sensors in a networked control system when some of the sensors are corrupted by an adversary. An operator with access to offline input-output attack-free trajectories of the plant is considered. Then, a data-driven algorithm is proposed to identify the attack-free sensors when the plant is controlled online. We also provide necessary conditions, based on the properties of the plant, under which the algorithm is feasible. An extension of the algorithm is presented to identify the sensors completely online against certain classes of attacks. The efficacy of our algorithm is depicted through numerical examples.
\end{abstract}

\section{Introduction}
The security of Networked Control Systems (NCS) has received considerable research attention in the past decade \citep{dibaji2019systems} due to the growing number of cyber-attacks. Many approaches have been proposed in the literature to design secure NCS against such attacks. Some of the approaches include (i) attack detection \citep{giraldo2018survey,li2023attack} and identification \citep{ameli2018attack,sakhnini2021physical}, (ii) secure state estimation \citep{fawzi2014secure,chong2015observability}, (iii) resilient controller design \citep{hashemi2019co}, and many more. 

This paper considers the problem of identifying attack-free sensors when some of the sensors are corrupted by attacks. In particular, this paper considers a Discrete-Time (DT) Linear Time-Invariant (LTI) plant whose $N$ sensor outputs are transmitted over the network. The adversary can manipulate up to $M$ sensor channels. The operator has access to offline input-output trajectories (say $\mathcal{I}_T$) of the plant. Here, the data $\mathcal{I}_T$ is assumed to be collected (in open-loop) before the plant is controlled online and thus is attack-free. To analyze the worst-case, we assume that the operator does not have the model of the plant. Then, the paper focuses on the following research problem.
\begin{problem}\label{problem}
Consider an LTI DT plant where the sensors are under potentially unbounded attack. Using only the offline attack-free input-output trajectories of the LTI plant, what are the conditions under which the attack-free sensors can be identified when the plant is controlled online? $\hfill \triangleleft$
\end{problem}

Once the operator identifies the attack-free sensors, techniques from the literature can be used to design stabilizing controllers or state estimators. There are several works in the literature that focus on the problem of designing stabilizing controllers from data without constructing an explicit state space model. For instance, \citet{de2019formulas} describes a data-driven approach for designing stabilizing controllers for LTI models using input-state data and ARX (autoregressive model with exogenous input) model with input-output data. Building on \citet{de2019formulas}, the work \citet{bisoffi2022data} describes an approach for designing stabilizing controllers using input and noisy state data. Recent work by \citet{steentjes2021data} describes a design approach for ARX systems from noisy input-output measurements using covariance bounds on the noise. The above works cannot be directly extended to solve Problem \ref{problem} since they assume certain statistical properties about the noise signals which are information we do not assume about attacks in this paper. 

Furthermore, the paper \citet{turan2021data} develops an algorithm to reconstruct the states in the presence of unknown inputs. The article \citet{van2020data} discusses the necessary conditions to reconstruct the states from output data. A summary of the results on direct data-driven control can be found in \citet{van2023informativity,markovsky2022data}. 

Albeit the above-mentioned works, to the best of the authors' knowledge, there are very few papers that examine the attack scenario through the data-driven framework for LTI models \citep{russo2021poisoning,russo2023analysis}. Also, the papers \citet{russo2021poisoning,russo2023analysis} design optimal attack policies on data-driven control methods but do not propose any defense/mitigation strategies against them. Thus, as a first step toward studying NCS under sensor attacks using the data-driven framework, this paper examines Problem \ref{problem} and presents the following contribution. 
\begin{enumerate}
    \item A data-driven algorithm is proposed to identify the attack-free sensors. 
    \item Necessary conditions are provided under which the algorithm returns attack-free sensors. 
    \item The proposed algorithm is extended for online identification (no access to attack-free trajectories) against certain classes of attacks (delay attacks and replay attacks). 
\end{enumerate}
Finally, as advocated in \citet{10317633}, the direct data-driven control (designing controllers from data) has some merits compared to the indirect data-driven approach (first identifying the model and then designing a control policy). The results presented in this paper act as stepping stones toward direct data-driven control in the presence of sensor attacks. 

\textit{Notation:} In this paper, $\mathbb{R}, \mathbb{C}$, and $\mathbb{Z}$ represent the set of real numbers, complex numbers, and integers respectively. 
An identity matrix of size $m \times m, m > 0$ is denoted by $I_m$. A zero matrix of size $n \times m, n >0,m >0$ is denoted by $0_{n \times m}$. Let $x: \mathbb{Z} \to \mathbb{R}^n$ be a discrete-time signal with $x[k]$ as the value of the signal $x$ at the time step $k$. The Hankel matrix associated with $x$ is denoted as 
\begin{equation}\label{Hankel_x}
X_{i,t,N} = \begin{bmatrix}
        x[i] & x[i+1] & \dots & x[i+N-1]\\
        x[i+1] & x[i+2] & \dots & x[i+N]\\   
        \vdots & \vdots & \ddots & \vdots\\
        x[i+t-1] & x[i+t] & \dots & x[i+t-1+N]
    \end{bmatrix}
\end{equation}
where the first subscript of $X$ denotes the time at which the first sample of the signal is taken, the second one the number of samples per column, and the last one the number of signal samples per row.  If the second subscript $t = 1$, the Hankel matrix is denoted by $X_{i,N}$. The notation $x_{[0,T-1]}$ denotes the vectorized, time-restricted signal $x$ which takes the following expression $x_{[0,T-1]} = \begin{bmatrix}x[0] & x[1] & \dots x[T-1]\end{bmatrix}$. The signal $x_{[0,T-1]}$ is defined to be persistently exciting of order $L$ if the matrix $X_{0,L,T-L+1}$ has full rank $Ln$. The $2$-norm of the vector $x\in \mathbb{R}^n$ is denoted as $||x||_2 = \sqrt{x^Tx}$. The cardinality of a vector $x\in \mathbb{R}^n$ is denoted as $|x|$.
\vspace{-10pt}
\section{Problem Description}\label{sec:PF}
\begin{figure}
    \centering
    \includegraphics[scale=0.4]{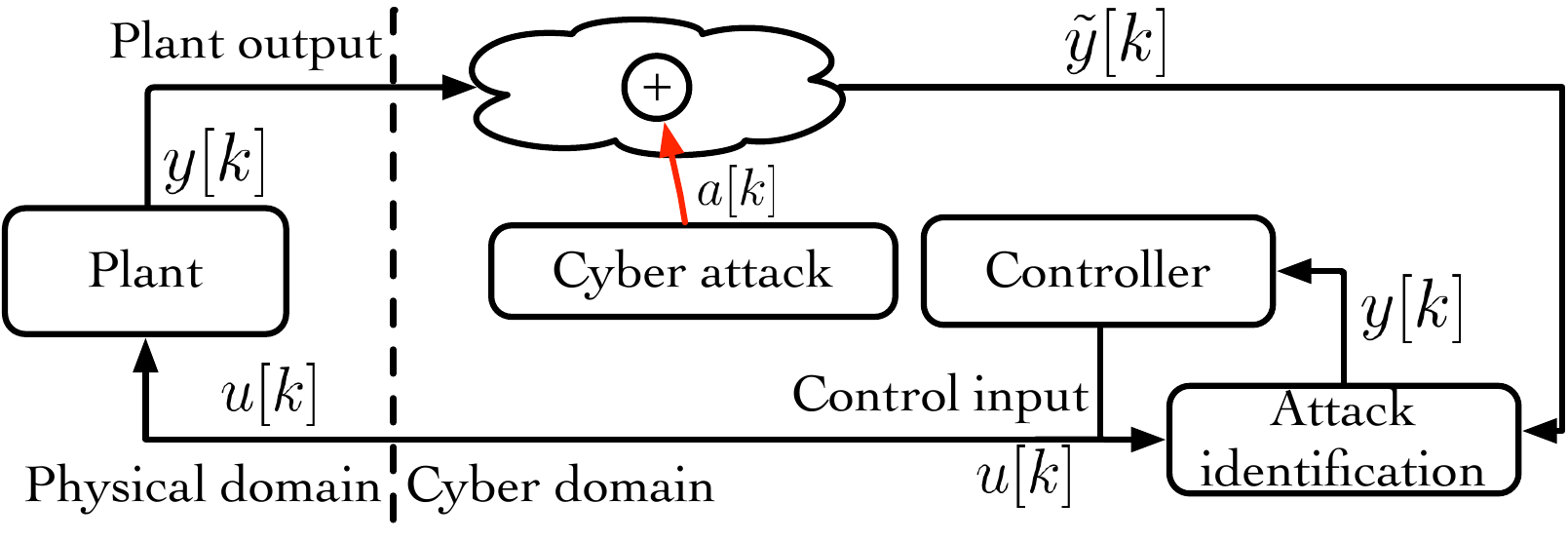}
    \caption{Networked control system under sensor attacks.}
    \label{fig:NCS}
    \vspace{-15pt}
\end{figure}

Consider a controllable LTI State-space (SS) model of the plant ($\mathcal{P}$) described as:
\begin{equation}\label{P}
\mathcal{P}: \qquad
    \begin{aligned}
x[k+1]&=Ax[k]+Bu[k], \quad 
y_i[k]= C_ix[k],\;i \in \{1,\dots,N\},
\end{aligned}
\end{equation}
where the state of the plant is represented by $x[k] \in \mathbb{R}^n$, the control input applied to the plant is represented by $u[k] \in \mathbb{R}^m$, the output of the noise-free sensor $i, i \in \{1,\dots,N\}$ is represented by $y_i[k] \in \mathbb{R}$ and the matrices $A, B,$ and $C_i$ are of appropriate dimensions. Without loss of generality, we assume that each sensor $i$ is of unit dimension, and the following notation is used for simplicity
\begin{equation}
y[k] =Cx[k], \quad y[k]= \begin{bmatrix}
        y_1[k]^T & y_2[k]^T & \dots & y_N[k]^T
    \end{bmatrix}^T, \quad
    C=\begin{bmatrix}
        C_1^T & C_2^T & \dots & C_N^T
    \end{bmatrix}^T.
\end{equation}
As shown in Figure.~\ref{fig:NCS}, the sensor measurements are transmitted over the network for control and monitoring purposes. Sensor measurements which are transmitted over a network are prone to cyber-attacks, where the adversary may manipulate the transmitted sensor measurements with malicious intent. We assume that the adversary can manipulate not more than $M$ out of $N$ sensors.
\begin{assumption} \label{assum:M-attacks}
The adversary can compromise up to $M$ out of $N$ sensor measurements $y_i$ for $i\in\{1,2,\dots,N\}$.$\hfill \triangleleft$
\end{assumption}

Here, the value of $M$ represents an upper bound on the number of sensors that the operator believes are susceptible to attack. For instance, it can represent the sensor channels for which the encryption key has not been updated recently. 
\subsection{Attack scenario}
As depicted in Figure.~\ref{fig:NCS}, an adversary injects malicious data into the sensors and is represented by 
$\Tilde{y}[k]=y[k]+a[k], \quad a[k]=\begin{bmatrix}
    a_1^T[k] & a_2^T[k] &  \dots & a_N^T[k] 
\end{bmatrix}^T,$
where $\Tilde y$ represents the attacked sensor signal received by the operator, and $a_i$ represents the attack signal on sensor channel $i$. In particular, three attack strategies are considered in this paper and are explained next. 
\begin{enumerate}
\item \textit{Sensor data-injection attacks} are of arbitrary magnitudes, but only corrupt at most $M$ sensors.
\item \textit{Sensor-network delay attacks} introduces arbitrary but bounded time delays in at most $M$ sensors.
\item \textit{Sensor replay attacks} records sensor values and replays them in at most $M$ sensors.
\end{enumerate}
Next, the main objective of this paper is stated.
\subsection{Problem formulation and approach}
As shown in Figure.~\ref{fig:NCS}, consider an operator who has a poor knowledge of the plant, i.e., the system matrices $A, B$, and $C$ are unknown. However, the operator has access to the input $u$ and (potentially corrupted) output $\tilde{y}$. The aim of the operator (and this paper) is as follows: when the plant is controlled/monitored online, stabilize the plant without model knowledge (matrices $A$, $B$ and $C$ are unknown) using input and output trajectories $u$ and $\tilde{y}$. Our approach to this problem is to first learn a model of the plant $\mathcal{P}$ offline (or open-loop) using input $u$ and output $y$ trajectories, compactly written as $\mathcal{I}_{T}$ of length $T$ (where $T$ is sufficiently large), i.e., $\mathcal{I}_{T}=\{u_{[0,T-1]},{y}_{[0,T-1]}\}.$
We call such a model the data-driven open-loop representation of the plant. The process of learning the model offline is necessary to obtain a representation that is free from sensor attack, which would then allow us to identify the sensors when the plant goes online. In this paper, we use the data-driven model to identify the sensors which have been corrupted with two classes of well-known attacks, namely, data injection attacks in Section \ref{sec:identify} and for delay and replay attacks in Section \ref{sec:online}.

\section{Model-free identification of data-injection attacks}\label{sec:identify}
We first derive a data-driven representation of the plant \eqref{P} offline, which gives us a data-driven open loop model that explains the input and output trajectories of length $T$, $\mathcal{I}_{T}$. 

\subsection{Learning the data-driven open-loop (offline) model}
Due to Assumption \ref{assum:M-attacks}, we know that at most $M$ sensors can be compromised, but we do not know which sensors have been compromised. Hence, our approach is to learn the open-loop models of the plant with every subset of $N-M$ sensors. In other words, we will learn a total of $(N-M)$-combinations of data-driven open-loop models from the data $\mathcal{I}_{T}$. To this end, let the plant \eqref{P} with output $z_j[k]$ for $j \in \left\{1,\dots,{N \choose N-M}\right\}$, be described as follows
\begin{equation}\label{eq:SS_con_form}
x[k+1]=Ax[k]+Bu[k], \quad z_j[k]=C_{{\mathcal{J}}_{j}}x[k], \; k\in\mathbb{Z},
\end{equation}
where ${\mathcal{J}}_{j}\subset \{1,\dots,N\}$ with cardinality $N-M$ and $C_{{\mathcal{J}}_{j}}$ is a matrix obtained by stacking all the matrices $C_i$ for $i\in {\mathcal{J}}_{j}$. Note that \eqref{P} and \eqref{eq:SS_con_form} are equivalent since their input and output trajectories match. Necessarily, we assume the following 
\begin{assumption}[$N-M$-sensor observability]\label{assum:q_sen_obsv}
For every subset $\mathcal{J}\subset \{1\dots,N\}$ with cardinality $N-M$, the tuple $(A,C_{\mathcal{J}})$ is observable where $C_{\mathcal{J}}$ is a matrix obtained by stacking the matrices $C_i,i\in \mathcal{J}$. $\hfill \triangleleft$
\end{assumption}
We say that the plant \eqref{P} is $(N-M)$-sensor observable when Assumption \ref{assum:q_sen_obsv} holds. Therefore, every $j\in\left\{1,\dots,{N \choose N-M}\right\}$ model described by \eqref{eq:SS_con_form} is observable. To construct a data-based representation of the plant \eqref{eq:SS_con_form} (or equivalently \eqref{P}) using $\mathcal{I}_{T}$, define 
\begin{equation}\label{data_matrices0}
\hat{X}_{j,n,T} \triangleq  \begin{bmatrix}
        Z_{j,0,n,T}\\
        \hline 
        U_{0,n,T}\\
    \end{bmatrix},\; \mathcal{X}_{j}[k]^T \triangleq
\setlength{\dashlinegap}{0.5pt}
\begin{bmatrix}[ccc:ccc] z_{j}^T[k-n]&...&z_j^T[k-1]& u^T[k-n]&...& u^T[k-1]\end{bmatrix}.
\end{equation}
where $Z_{j,0,n,T}$ is defined similar to \eqref{Hankel_x} but with signals $z_j$ instead of $x$ as in \eqref{Hankel_x}. Then the operator constructs the following matrices from the data. 
\begin{align}
    U_{n,T}&=\begin{bmatrix}
        u[n] & u[n+1]& \dots & u[n+T-1]
    \end{bmatrix} \label{data_matrices}\\
    \hat{X}_{j,n,T}&=\begin{bmatrix}
        \mathcal{X}_{j}[n] & \mathcal{X}_{j}[n+1]& \dots & \mathcal{X}_{j}[n+T-1]
    \end{bmatrix} \label{data_matrices1}\\
    \hat{X}_{j,n+1,T}&=\begin{bmatrix}
        \mathcal{X}_{j}[n+1] & \mathcal{X}_{j}[n+2]& \dots & \mathcal{X}_{j}[n+T]
    \end{bmatrix} \label{data_matrices2}
\end{align}
Note that the state $\mathcal{X}_j$ in contains the history of past inputs and past outputs of length $n$. Thus there is an offset of length $n$ whilst constructing the data matrices \eqref{data_matrices}. Now the following result is obtained using \citet[Theorem 1, Lemma 2]{de2019formulas}.
\begin{lemma}\label{lem:data_OL}
For any given $j \in \left\{1,\dots,{N \choose N-M}\right\}$, let
\begin{equation}\label{eq:PE}
    \text{rank}\begin{bmatrix}
        U_{n,T}\\\hat{X}_{j,n,T}
    \end{bmatrix}=m(n+1)+(N-M)n,
\end{equation}
with $T\geq (m+1)\tilde{n}$ where $\tilde{n}\triangleq (m+N-M)n+1$. Then \eqref{eq:SS_con_form} has the equivalent data-driven representation in \eqref{data:OL}, where $\ssymbol{2}$ denotes the right inverse.
\begin{equation}\label{data:OL}
\mathcal{X}_j[k+1]=\Lambda_j\begin{bmatrix}
    u[k]\\ \mathcal{X}_j[k]
\end{bmatrix},\;\text{where}\;\Lambda_j \triangleq \hat{X}_{j,n+1,T}
\begin{bmatrix}
U_{n,T}\\
\hat{X}_{j,n,T}
\end{bmatrix}\ssymbol{2}. \quad \square
\end{equation}
\end{lemma}

The proof of Lemma \ref{lem:data_OL} is in Appendix \ref{app3}. Lemma \ref{lem:data_OL} states that when the rank condition \eqref{eq:PE} is satisfied by the data $\mathcal{I}_T$, then \eqref{eq:SS_con_form} has an equivalent data-driven representation \eqref{data:OL}. An equivalent and simpler method to check if \eqref{eq:PE} holds was given in \citet[Lemma 1, and (81)]{de2019formulas} which requires checking if the input $u$ is persistently exciting of a certain order. This equivalent result is next recalled by including a remark that the observability of the data-generating matrices is a necessary condition for \eqref{eq:PE} to hold: highlighting the necessity of Assumption \ref{assum:q_sen_obsv}.
\begin{lemma}\label{lem:rank_obsv}
Let the input $u_{[n,n+T-1]}$ with $T\geq (m+1)\tilde{n}$ be persistently exciting of order $\tilde{n}$ where $\tilde{n}=(m+N-M)n+1$. Then for a given $j \in \left\{1,\dots,{N \choose N-M}\right\}$, \eqref{eq:PE} cannot hold in the absence of attacks, if the tuple $(A,C_{{\mathcal{J}}_{j}})$ is not observable and $n < N-M+1$. $\hfill \square$
\end{lemma}
The proof of Lemma \ref{lem:rank_obsv} is given in Appendix \ref{app2}. Using this data-driven representation derived in Lemma \ref{lem:data_OL}, the algorithm to identify the attack-free sensors is described next.
\subsection{Identification of data injection attacks}\label{sec22}
This section aims to identify the attack-free sensors online, using the open-loop representation in \eqref{data:OL}. To this end, consider the open-loop (offline) data consisting of inputs $u$ and measurements $\left\{z_1,z_2,\dots,z_{{N \choose N-M}}\right\}$. Using the data, construct the data matrices in \eqref{data_matrices}, \eqref{data_matrices1}, \eqref{data_matrices2}, and $\Lambda_j$ for $j \in \{1,2,\dots, {N\choose N-M} \}$. Now, the matrix $\Lambda_j$ is the model for the operator, and all input $u$ and attack-free output $y$ trajectories that satisfy \eqref{P} also satisfy \eqref{data:OL}. Now, consider the online measurements $\tilde{z}_{j}=\tilde{y}_{\mathcal{J}_{j}},\;j\in\left\{1,2,\dots, {N\choose N-M} \right\},$
where $\tilde{y}_{\mathcal{J}_{j}}$ denotes the stacking of $\tilde{y}_{i}$ for $i\in\mathcal{J}_{j}$. Note that the online measurements $\tilde{z}_{j}$ can also be written as $\tilde{z}_{j}=z_j + a_{{\mathcal{J}_{j}}}$ and are potentially corrupted. By construction, $\exists\;\tilde{z}_j, j\in \{1,2,\dots, {N\choose N-M} \}$ that is attack-free. Then, for a given input $u$, the data-driven open-loop (offline) model given in \eqref{data:OL} explains the attack-free sensor measurements $\tilde{z}_{j}$. Motivated by the above argument, at any given time instant $k \in \mathbb{Z}$, let us apply a new test input signal (say $\underline{u} \neq 0$) of length $1$, collect the corresponding outputs $\underline{y}$ and construct the new data set 
\begin{equation}\label{eq:test_data}
\underline{\mathcal{I}}_{T_1} \triangleq \left\{\underline{u}{[k]},\underline{y}{[k+1]} \right\}.
\end{equation}
As before, let us construct the data matrices in \eqref{data_matrices} using the test data $\underline{\mathcal{I}}_{T_1}$. Let us represent the data matrices as $\underline{U}_{k,1}, \underline{\hat{X}}_{j,k,1},$ and $\underline{\hat{X}}_{j,k+1,1}$. Then, the attack-free sensors at any time instant $k\in\mathbb{Z}$ are given by 
\begin{equation}\label{eq:attack_cond}
    j^*[k+1] \in \underset{{j\in \left\{1,2,\dots, {N \choose N-M} \right\}}}{\arg\min} \Bigg|\Bigg| \underline{\hat{X}}_{j,k+1,1}-\Lambda_j\begin{bmatrix}
        \underline{U}_{k,1}\\
        \underline{\hat{X}}_{j,k,1}
    \end{bmatrix}\Bigg|\Bigg|_2.
\end{equation}
The procedure described above is depicted as an algorithm in Algorithm \ref{alg1}. Note that, for computing the norm in \eqref{eq:attack_cond}, we need access to $ \underline{\hat{X}}_{j,k,1}$  defined in \eqref{data_matrices0} which is the history of inputs and online measurements (of length $n$). At the beginning of the algorithm, $ \underline{\hat{X}}_{j,k,1}$ can be obtained by simply collecting the data in open loop. After which, $ \underline{\hat{X}}_{j,k,1}$ can be constructed in a moving horizon fashion. 
Combining the above arguments, the following result is stated.
\begin{prop}\label{prop1}
Consider the plant \eqref{P} under Assumption \ref{assum:M-attacks} and \ref{assum:q_sen_obsv}. Given that the attack starts after $n$ time steps, then ${j^*}$ obtained from Algorithm \ref{alg1} is the vector of attack-free sensors.
$\hfill \square$
\end{prop}
The proof of Proposition \ref{prop1} is given in Appendix \ref{app5}. The result stated in Proposition \ref{prop1} relies on the fact that the plant \eqref{P} has an equivalent data-driven representation in \eqref{data:OL}. As stated in Lemma \ref{lem:data_OL}, the observability of the data generating system is a necessary condition for the rank condition \eqref{eq:PE} to hold, which is in turn necessary for the existence of the data-driven representation \eqref{data:OL} as stated in Lemma \ref{lem:data_OL}. Thus, Assumption \ref{assum:q_sen_obsv} is a necessary condition for the data-driven representation \eqref{data:OL} to be equivalent to \eqref{P}, and for Algorithm \ref{alg1} to work. The next section provides a discussion on the identification of attack-free sensors without access to $\mathcal{I}_T$ in the presence of sensor delay attacks.
\begin{algorithm}
\caption{Algorithm to identify data injection attack-free sensors}\label{alg1}
\begin{algorithmic}[1]
\State (Offline) Using $\mathcal{I}_T$, construct $\Lambda_j, \; \forall j \in \left\{1,2,\dots, {N \choose N-M} \right\}$ from Lemma \ref{lem:data_OL}. Set $k=n$.
\State (Online) Apply random inputs of length $n$, collect online measurements and construct $\underline{\hat{X}}_{j,k,1}$
\State (Online) Apply \textit{test} input data of length $1$, collect the corresponding output data, and construct the matrices $\underline{U}_{k,1}, \underline{\hat{X}}_{j,k,1},$ and $\underline{\hat{X}}_{j,k+1,1}$ for all $j \in \left\{1,2,\dots, {N \choose N-M} \right\}$. 
\State (Online) Determine $j^*[k+1]$ from \eqref{eq:attack_cond}.
\State If no sensors are under attack, i.e.: the R.H.S of \eqref{eq:attack_cond} is $\left\{1,2,\dots, {N \choose N-M} \right\}$, then set $\qquad \qquad k=k+1, \underline{\hat{X}}_{j,k,1}=\underline{\hat{X}}_{j,k+1,1}$, and go to step $3$
\end{algorithmic}
\end{algorithm}
\vspace{-10pt}
\section{Online identification of attacks without using attack-free data }\label{sec:online}
In the previous section, we discussed the identification of attacks when we have access to offline data of fixed length. However, in some scenarios, offline data might not be available. Thus in this section, we discuss cases in which attack identification can be performed without $\mathcal{I}_T$ before which we establish the following assumption. 
\begin{assumption}\label{ass_eq}
The plant \eqref{P} is at equilibrium $x[0]=0$ before the attack commences. $\hfill \triangleleft$
\end{assumption}
Assumption \ref{ass_eq} does not introduce any loss of generality due to the following reasons. In reality, the plant is usually stabilized locally with a feedback controller (for safety), and a network controller is used to improve performance, change set points, etc. This approach of having two controllers is adopted in the literature \citep{han2018stability} to avoid stability issues due to packet dropouts, delays, bandwidth limitations, etc \citep{hu2007stability}. Thus, the stable plant (along with the local controller) reaches an equilibrium from any non-zero initial condition satisfying Assumption \ref{ass_eq}. Such assumptions are also common in the literature \citep{wang2020neural,nonhoff2022online}.

\subsection{Replay attacks}
In this section, we consider the case of replay attacks \cite{teixeira2015secure}. We consider an adversary who has recorded the sensor measurements which are constants. These constants can correspond to the equilibrium point (Assumption \ref{ass_eq}) or any other operating point. For instance, if the local controller is designed for reference tracking, the outputs might correspond to the constant reference trajectory. We represent replay attacks as 
\begin{equation}\label{replay}
    \tilde{y}_i[k] = c_i, \forall i \in \{1,\dots,N\}, \forall k \in \mathbb{Z},
\end{equation}
where $c_i$ is a constant unknown to the operator. Another approach to motivate attacks in \eqref{replay} is as follows. Under Assumption \ref{assum:M-attacks}, let us consider the plant is under constant bias injection attacks \citep{teixeira2015secure,farraj2017impact,jin2022distributed}. Then, the attack can be modeled as \eqref{replay} where the $c_i$ denotes the bias injected into sensor $i$. The effect of bias injection attacks (also when the bias is slowly increasing) on experimental setups was also recently studied \citep{kedar2023evaluation}. Next, we are interested in detecting the sensors that are replay attack-free.

Motivated by the discussion around Algorithm \ref{alg1}, let us apply test input signals (say $\underline{u}$) of length $T_1$, collect the corresponding outputs $\underline{y}$ and construct the data set $\underline{\mathcal{I}}_{T_1}$ as in \eqref{eq:test_data}. As before, let us construct the data matrices in \eqref{data_matrices} using the test data $\underline{\mathcal{I}}_{T_1}$ in \eqref{eq:test_data}. Let us represent the data matrices as $\underline{U}_{n,T_1}, \underline{\hat{X}}_{j,n,T_1},$ and $\underline{\hat{X}}_{j,n+1,T_1}$. Then, we present the following result

\begin{lemma}\label{lem:replay}
Consider the plant \eqref{P} under replay attacks \eqref{replay}, Assumption \ref{assum:M-attacks}, and Assumption \ref{ass_eq}. Let the input $\underline{u}_{[n, n+T_1-1]}$ with $T_1\geq (m+1)\tilde{n}$ be persistently exciting of order $\tilde{n}$ where $\tilde{n}\triangleq (m+N-M)n+1$. Then, if the sensor pair $j \in \left\{1, 2, \dots {N \choose N-M} \right\}$ is under replay attack, the following rank condition 
\begin{equation}\label{rank:replay}
    \text{rank}\begin{bmatrix}
        \underline{U}_{n,T_1}\\ \underline{\hat{X}}_{j,n,T_1}
    \end{bmatrix}=m(n+1)+(N-M)n,
\end{equation}
cannot hold. $\hfill \square$
\end{lemma}
The proof of Lemma \ref{lem:replay} is in Appendix \ref{App6}. Lemma \ref{lem:replay} says that, if the sensors are under replay attack, the rank condition in \eqref{rank:replay} can be used as a test to detect replay attacks since it is satisfied for persistently exciting inputs as stated in Lemma \ref{lem:rank_obsv}. We sketch the procedure above in Algorithm \ref{alg:replay} and the corresponding result in Proposition \ref{prop:replay}.
\begin{proposition}\label{prop:replay}
Consider the plant \eqref{P} under Assumption \ref{assum:M-attacks}, \ref{assum:q_sen_obsv} and \ref{ass_eq}. Then ${j^*}$ obtained from Algorithm \ref{alg:replay} is the vector of replay attack-free sensors. $\hfill \square$
\end{proposition}
The proof of Proposition \ref{prop:replay} is given in Appendix \ref{App7}. Next, we briefly discuss the online identification of delay attacks.
\begin{algorithm}
\caption{Algorithm to identify replay attack-free sensors}\label{alg:replay}
\begin{algorithmic}[1]
\State Apply persistently exciting inputs $\underline{u}$ of length $T_1$. 
\State Collect the online measurements and construct $\underline{U}_{n,T_1}, \underline{\hat{X}}_{j,n,T_1}, \forall j \in \{1, 2,\dots, {N \choose N-M}\}$.
\State Determine $\; j^* \in \arg \max_{j \in \{ 1, 2,\dots, {N \choose N-M} \}} \left\{ \text{rank}\begin{bmatrix}
        \underline{U}_{n,T_1}\\ \underline{\hat{X}}_{j,n,T_1}
    \end{bmatrix} \right\}$
\end{algorithmic}
\end{algorithm}
\vspace{-15pt}
\subsection{Network delay attacks}
As mentioned in the introduction, there are different kinds of attacks studied in the literature which include stealthy data-injection attacks, replay attacks, zero-dynamics attacks, covert attacks, and many more \citep{teixeira2015secure}. One of the attacks which have not been studied intensively is time delay attacks \citep{bianchin2018time,wigren2023line}. Delay attacks are shown to destabilize power grids \citep{korkmaz2016ics}, however, their detection schemes are not well developed. In this paper, we first represent delay attacks on each $i$-th sensor as follows 
\begin{equation}
    \tilde{y}_i[k] = y_i[k-\tau_i], \tau_i \geq 0, \tau_i \in \mathbb{Z}, \forall i \in \{1,\dots,N\},
\end{equation}
where $\tilde{y}_i[k]$ is the delayed measurement received by the operator. Next, we attempt to identify the sensors which are free of delay attacks. However, we consider single-input systems for simplicity. We start the discussion by defining the relative degree of an LTI system.
\begin{definition}\label{def:rd}
The relative degree $(r_j)$ of the system $\Sigma_j=(A,B,C_j,0), j\in\{1,\dots,N\}$ is defined as $ r_j \triangleq 1+\min \{ i\in\mathbb{Z} \;| C_jA^iB \neq 0, i \geq 0 \}. \hfill \triangleleft$
\end{definition}
The relative degree of $\Sigma_j$ denotes the amount of time delay before the input $u[k]$ appears in its output $y_j[k]$. For instance, consider a system with a feed-through term $D \neq 0$, then the input $u[k]$ immediately appears in the output making the relative degree of $\Sigma$ zero. Next, we establish the following assumptions. 
\begin{assumption}\label{ass2}
We do not have offline data $\mathcal{I}_T$, and we know the value of $r_j$, $\forall \; j\in\{1,\dots,N\}. \hfill \triangleleft$
\end{assumption}
In other words, we assume that we have some degree of knowledge about the plant in terms of the relative degree (instead of attack-free data). For instance, the relative degree can be obtained from the physics of the plant. That is, if the operator knows the physics of the plant but does not know the exact parameters of the plant (as considered in \cite{de2023learning}), then the operator can infer relative degrees. With this knowledge of the relative degrees, under Assumptions \ref{assum:q_sen_obsv}, \ref{ass_eq} and \ref{ass2}, we next describe an algorithm for identification of sensors which are free from delay attacks.

Let us apply a non-zero input signal of unit length (approximate impulse) and collect the corresponding outputs. Since the input is non-zero, and the plant is at equilibrium before the attack, the outputs corresponding to the sensor $y_j$ should be non-zero after $r_j$ time steps. If the output of the sensor $y_j$ is zero after $r_j$ time steps, then the corresponding sensor is under attack. Then, the attack-free sensors (denoted by $j^* \in \{1, 2,\dots,N\}$) at any time instant are given by 
\begin{align}\label{eq:att_delay}
    j^* &\in \arg\min_{j\in \{1, 2,\dots, N\}} \left\{ \sigma_j - r_j\right\}, \quad \sigma_j \in \min \{k \in \mathbb{Z} \vert Y_{j,0,T}[1,k] \neq 0\},
\end{align}
where $Y_{j,0,T}$ is defined similar to \eqref{Hankel_x} but with the output signals $y$ instead of $x$ as in \eqref{Hankel_x}, and $Y_{j,0,T}[1,k]$ denotes the element of the matrix $Y_{j,0,T}$ in row $1$ and column $k$. The idea behind \eqref{eq:att_delay} is that the equation on the right determines the first column of $Y_j$ which is non-zero. Ideally, when there is no attack, the first non-zero column of $Y_j$ would be $r_j$. The procedure described above is depicted as an algorithm in Algorithm \ref{alg:delay}. We also state the corresponding result in Proposition \ref{prop:delay}.
\begin{algorithm}
\caption{Algorithm to identify delay attack-free sensors}\label{alg:delay}
\begin{algorithmic}[1]
\State Apply \textit{non-zero} input vector of (unit) length $1$.
\State Collect the outputs and construct the matrices $Y_{j,0,T}, \forall j \in \{1,\dots, {N}\}$ where $T \geq  \max_j{r_j}$.
\State Determine $j^*$ from \eqref{eq:att_delay}
\end{algorithmic}
\end{algorithm}
\vspace{-15pt}
\begin{proposition}\label{prop:delay}
Consider the plant \eqref{P} under Assumption \ref{ass_eq}, and let at least one of the sensors be attack-free. Then ${j^*}$ obtained from Algorithm \ref{alg:delay} is the vector of delay attack-free sensors. $\hfill \square$
\end{proposition}
The proof of Proposition \ref{prop:delay} is given in Appendix \ref{App8}. In this section, akin to the experimental design techniques for data-driven learning \citep{de2021designing}, we proposed input design techniques for attack detection. Next, we show the efficacy of our algorithms through numerical examples.
\section{Numerical Example}\label{sec:NE}
\begin{wrapfigure}{r}{10cm}
    \centering
    \includegraphics[scale=0.4]{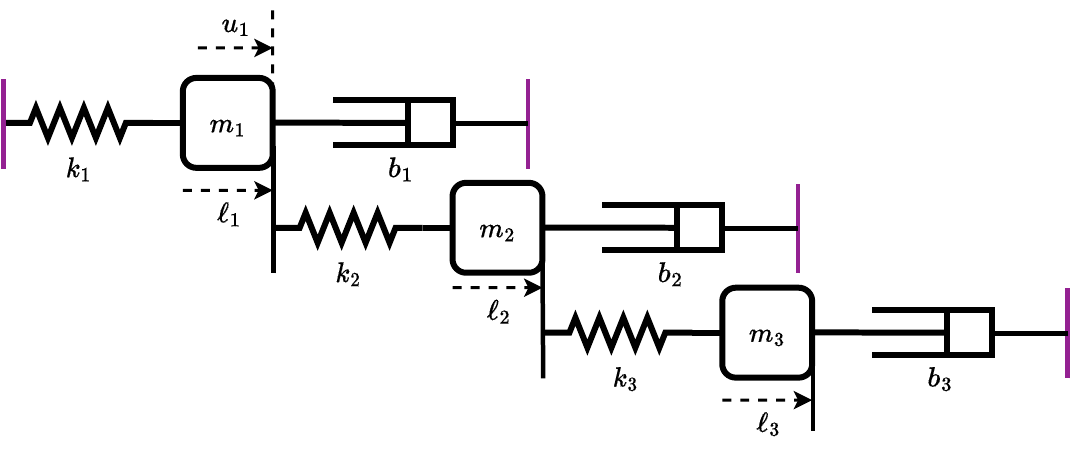}
    \caption{Three interconnected mass spring damper model. The lines in purple denote fixed supports.}
    \label{fig:MSD}
\end{wrapfigure}
Consider the interconnected mass spring damper model in Figure.~\ref{fig:MSD} where the displacement of mass $m_i, i=\{1,2,3\}$ is represented by $\ell_i$. The input is the force $(u)$ applied to mass $m_1$. The outputs are the displacement ($\ell_2$) of mass $m_2$, the displacement ($\ell_3$) of mass $m_3$, and the velocity ($\dot{\ell}_3$) of  mass $m_3$. The parameters of the model are $k1=$
$k_1=2$ \mbox{N/m}, $m_1=1$ \mbox{Kg}, $b_1=3$ \mbox{Ns/m}, $k_2=3$ \mbox{N/m}, $m_2=2$ \mbox{Kg}, $b_2=4$ \mbox{Ns/m}, $k_3=1$ \mbox{N/m}, $m_3=10$ \mbox{Kg}, and $b_3=2$ \mbox{Ns/m}.

Using Newton's laws of motion, and defining the state vector as 
$x = \begin{bmatrix}
    l_1 & \dot{l}_1 & l_2 & \dot{l}_2 & l_3 & \dot{l}_3
\end{bmatrix}$
we arrive at the continuous-time (CT) state-space model for the interconnected model in \eqref{eq:SS_NE}.
\begin{equation}
    \dot{x} = \begin{bmatrix}
    0 & 1 & 0 & 0 & 0 & 0\\
    \frac{-k_1}{m_1} & \frac{-b_1}{m_1} & 0 & 0 & 0 & 0\\
    0 & 0 & 0 & 1 & 0 & 0\\
    \frac{1}{m_2} & 0 & \frac{-k_2}{m_2} & \frac{-b_2}{m_2} & 0 & 0\\
    0 & 0 & 0 & 0 & 0 & 1\\
    0 & 0 & \frac{1}{m_3} & 0 & \frac{-k_3}{m_3} & \frac{-b_3}{m_3}
    \end{bmatrix}x+\begin{bmatrix}
        0\\
        \frac{1}{m_1}\\
        0\\
        0\\
        0\\
        0
    \end{bmatrix}u,\;y =\begin{bmatrix}
        0 & 0 & 1 & 0 & 0 & 0\\
    0 & 0 & 0 & 0 & 1 & 0\\
    0 & 0 & 0 & 0 & 0 & 1
    \end{bmatrix}x\label{eq:SS_NE}
\end{equation}

\begin{figure}
\centering
\begin{minipage}{.5\textwidth}
  \centering
  \includegraphics[width=1\linewidth]{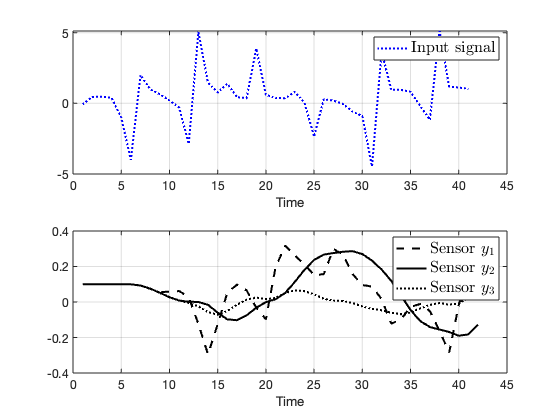}
  \captionof{figure}{\small{Attack-free \textit{offline} sensor data collected\\ to construct data matrices in \eqref{data_matrices} (bottom) and the \\corresponding input signal (top)}}
  \label{fig:sim}
\end{minipage}%
\begin{minipage}{.5\textwidth}
  \centering
  \includegraphics[width=1\linewidth]{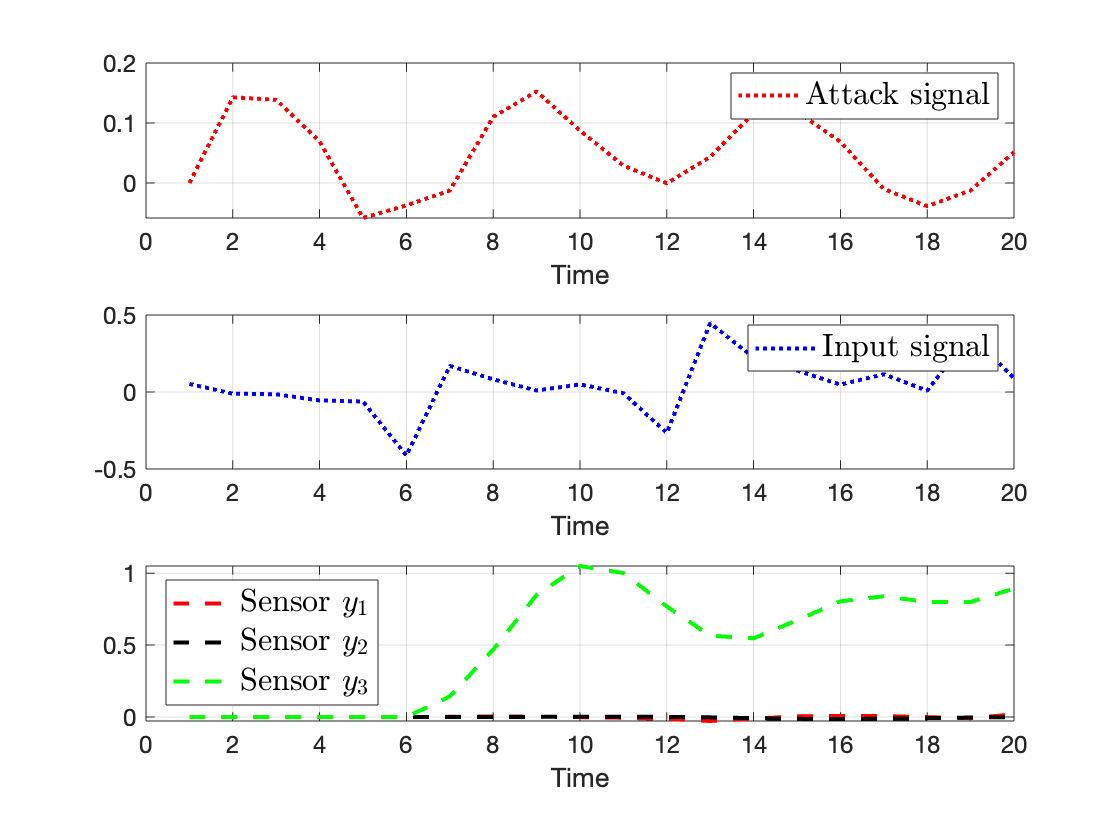}
  \captionof{figure}{\small{Attacked sensor \textit{test} data collected online for attack identification.}}
  \label{fig:sim2}
\end{minipage}
\end{figure}

We discretize the CT SS model with a sampling time of $T_s=1.3$ \mbox{s} and denote the DT SS matrices as $\Sigma = (A,B,C,0)$. The system $\Sigma$ is $2$-sensor observable (see Assumption \ref{assum:q_sen_obsv}). Thus, in this numerical example, we have $N=3$ (number of sensors), $2$-sensor observability, and $M=1$ ($Q=N-M$). Here $M$ represents the number of sensors that can be attacked. Then we have $\bar{\mathcal{J}}_{Q}=\{\{1,2\},\{2,3\},\{1,3\}\}$. Next, by following the results in Lemma \ref{lem:data_OL}, we set $T=41$. We apply input $u$ of length $T$, and collect the data ${y}$ and construct the matrices \eqref{data_matrices}, and $\Lambda_j$ for $j=\{1,2,3\}$, such that the rank condition \eqref{eq:PE} is satisfied. The collected input-output data is shown in Figure.~\ref{fig:sim}. 

\textit{Static attacked sensors:} Let us consider that sensor $3$ under attack. The attack injected into sensor $3$ is given in Figure.~\ref{fig:sim2} (top). To recall, we aim to identify the attack-free sensors using the attack-free open-loop representation. Next, we collect the \textit{test} data (Figure.~\ref{fig:sim2} (bottom)) and construct the test data matrices $\underline{U}_{k,T_1}, \underline{\hat{X}}_{j,k,1},$ and $\underline{\hat{X}}_{j,k+1,1}$ for all $j \in \{1,2,3\}$. The corresponding input signal is plotted in Figure.~\ref{fig:sim2} (middle). Using the constructed test data matrices, we determine the value of the norm in \eqref{eq:attack_cond} for $j=\{1,2,3\}$. The value of the norm is minimum (and zero) for sensor pairs $\{1,2\}$: denoting that they are attack-free. We were able to detect the attacked sensor in 2 time steps (since the value of the attack at the first time step is zero).

\textit{Network delay attacks:} For this example, we consider the following modified output matrix $C_m=0.1 \times C$. The relative degree of the sensors is $r_1=1,r_2=2,$ and $r_3=1$. We next consider an attacker introducing a network delay of $5$ samples into sensor $2$. We apply an input signal of the form $u[k]=0.1\delta[k]$ where $\delta[k]$ is a unit impulse. We see that the first non-zero input signal does not occur at the $2^{\text{nd}}$ time step. Thus, an attack is detected in sensor 2.

\textit{Replay attacks:} Let sensor 3 be under attack. The attack injected into sensor $3$ is a constant value $0.01$. We set $T=41$ to collect the \textit{test} data and construct the test data matrices in Lemma \ref{lem:replay}. Under replay attack, we observe that the matrix $\underline{\hat{X}}_{j,k+n,T_1}$ is rank deficient. Thus the rank test acts as a detection method against replay attacks.
\section{Conclusion}\label{sec:Con}
This paper studied the problem of attack-free sensor identification in a networked control systems when some of the sensors are corrupted by an adversary. An algorithm to identify the attack-free sensors using the offline attack-free data was proposed. The conditions under which the algorithm returns the attack-free sensors were discussed based on the observability of the data-generating matrices. A brief extension to identification without offline data was discussed. Future works include extending the proposed framework to include noise. 
\newpage
\acks{This work was conducted when Sribalaji. C. Anand was a visiting student at TU Eindhoven. Sribalaji C. Anand and Andr{\'e} M. H. Teixeira are supported by the Swedish Research Council under grant 2018-04396, the Swedish Foundation for Strategic Research. Sribalaji C. Anand was additionally supported by the Stenholm Wilgott travel scholarship.}
\bibliography{ref}

\begin{thebibliography}{33}
\providecommand{\natexlab}[1]{#1}
\providecommand{\url}[1]{\texttt{#1}}
\expandafter\ifx\csname urlstyle\endcsname\relax
  \providecommand{\doi}[1]{doi: #1}\else
  \providecommand{\doi}{doi: \begingroup \urlstyle{rm}\Url}\fi

\bibitem[Ameli et~al.(2018)Ameli, Hooshyar, El-Saadany, and Youssef]{ameli2018attack}
Amir Ameli, Ali Hooshyar, Ehab~F El-Saadany, and Amr~M Youssef.
\newblock Attack detection and identification for automatic generation control systems.
\newblock \emph{IEEE Transactions on Power Systems}, 33\penalty0 (5):\penalty0 4760--4774, 2018.

\bibitem[Antsaklis and Michel(1997)]{antsaklis1997linear}
Panos~J Antsaklis and Anthony~N Michel.
\newblock \emph{Linear systems}, volume~8.
\newblock Springer, 1997.

\bibitem[Bianchin and Pasqualetti(2018)]{bianchin2018time}
Gianluca Bianchin and Fabio Pasqualetti.
\newblock Time-delay attacks in network systems.
\newblock \emph{Cyber-Physical Systems Security}, pages 157--174, 2018.

\bibitem[Bisoffi et~al.(2022)Bisoffi, De~Persis, and Tesi]{bisoffi2022data}
Andrea Bisoffi, Claudio De~Persis, and Pietro Tesi.
\newblock Data-driven control via petersen’s lemma.
\newblock \emph{Automatica}, 145:\penalty0 110537, 2022.

\bibitem[Chong et~al.(2015)Chong, Wakaiki, and Hespanha]{chong2015observability}
Michelle~S Chong, Masashi Wakaiki, and Joao~P Hespanha.
\newblock Observability of linear systems under adversarial attacks.
\newblock In \emph{2015 American Control Conference (ACC)}, pages 2439--2444. IEEE, 2015.

\bibitem[De~Persis and Tesi(2019)]{de2019formulas}
Claudio De~Persis and Pietro Tesi.
\newblock Formulas for data-driven control: Stabilization, optimality, and robustness.
\newblock \emph{IEEE Transactions on Automatic Control}, 65\penalty0 (3):\penalty0 909--924, 2019.

\bibitem[De~Persis and Tesi(2021)]{de2021designing}
Claudio De~Persis and Pietro Tesi.
\newblock Designing experiments for data-driven control of nonlinear systems.
\newblock \emph{IFAC-PapersOnLine}, 54\penalty0 (9):\penalty0 285--290, 2021.

\bibitem[De~Persis et~al.(2023)De~Persis, Rotulo, and Tesi]{de2023learning}
Claudio De~Persis, Monica Rotulo, and Pietro Tesi.
\newblock Learning controllers from data via approximate nonlinearity cancellation.
\newblock \emph{IEEE Transactions on Automatic Control}, 2023.

\bibitem[Dibaji et~al.(2019)Dibaji, Pirani, Flamholz, Annaswamy, Johansson, and Chakrabortty]{dibaji2019systems}
Seyed~Mehran Dibaji, Mohammad Pirani, David~Bezalel Flamholz, Anuradha~M Annaswamy, Karl~Henrik Johansson, and Aranya Chakrabortty.
\newblock A systems and control perspective of {CPS} security.
\newblock \emph{Annual reviews in control}, 47:\penalty0 394--411, 2019.

\bibitem[Dörfler(2023)]{10317633}
Florian Dörfler.
\newblock Data-driven control: Part two of two: Hot take: Why not go with models?
\newblock \emph{IEEE Control Systems Magazine}, 43\penalty0 (6):\penalty0 27--31, 2023.

\bibitem[Farraj et~al.(2017)Farraj, Hammad, and Kundur]{farraj2017impact}
Abdallah Farraj, Eman Hammad, and Deepa Kundur.
\newblock On the impact of cyber attacks on data integrity in storage-based transient stability control.
\newblock \emph{IEEE Transactions on Industrial Informatics}, 13\penalty0 (6):\penalty0 3322--3333, 2017.

\bibitem[Fawzi et~al.(2014)Fawzi, Tabuada, and Diggavi]{fawzi2014secure}
Hamza Fawzi, Paulo Tabuada, and Suhas Diggavi.
\newblock Secure estimation and control for cyber-physical systems under adversarial attacks.
\newblock \emph{IEEE Trans. on Automatic control}, 59\penalty0 (6):\penalty0 1454--1467, 2014.

\bibitem[Giraldo et~al.(2018)Giraldo, Urbina, Cardenas, Valente, Faisal, Ruths, Tippenhauer, Sandberg, and Candell]{giraldo2018survey}
Jairo Giraldo, David Urbina, Alvaro Cardenas, Junia Valente, Mustafa Faisal, Justin Ruths, Nils~Ole Tippenhauer, Henrik Sandberg, and Richard Candell.
\newblock A survey of physics-based attack detection in cyber-physical systems.
\newblock \emph{ACM Computing Surveys (CSUR)}, 51\penalty0 (4):\penalty0 1--36, 2018.

\bibitem[Han et~al.(2018)Han, Tucci, Martinelli, Guerrero, and Ferrari-Trecate]{han2018stability}
Renke Han, Michele Tucci, Andrea Martinelli, Josep~M Guerrero, and Giancarlo Ferrari-Trecate.
\newblock Stability analysis of primary plug-and-play and secondary leader-based controllers for dc microgrid clusters.
\newblock \emph{IEEE Transactions on Power Systems}, 34\penalty0 (3):\penalty0 1780--1800, 2018.

\bibitem[Hashemi and Ruths(2022)]{hashemi2019co}
Navid Hashemi and Justin Ruths.
\newblock Co-design for resilience and performance.
\newblock \emph{IEEE Transactions on Control of Network Systems}, pages 1--12, 2022.
\newblock \doi{10.1109/TCNS.2022.3229774}.

\bibitem[Hu and Yan(2007)]{hu2007stability}
Shawn Hu and Wei-Yong Yan.
\newblock Stability robustness of networked control systems with respect to packet loss.
\newblock \emph{Automatica}, 43\penalty0 (7):\penalty0 1243--1248, 2007.

\bibitem[Jin et~al.(2022)Jin, Liu, Deng, and Cheng]{jin2022distributed}
Zexuan Jin, Mengxiang Liu, Ruilong Deng, and Peng Cheng.
\newblock Distributed data recovery against false data injection attacks in {DC} microgrids.
\newblock In \emph{2022 IEEE Intl. Conf. on Communications, Control, and Computing Technologies for Smart Grids (SmartGridComm)}, pages 265--270. IEEE, 2022.

\bibitem[Kedar Vadde~Hulgesh(2023)]{kedar2023evaluation}
Teja Kedar Vadde~Hulgesh.
\newblock Evaluation of cyber-attacks in networked control systems, 2023.

\bibitem[Korkmaz et~al.(2016)Korkmaz, Dolgikh, Davis, and Skormin]{korkmaz2016ics}
Emrah Korkmaz, Andrey Dolgikh, Matthew Davis, and Victor Skormin.
\newblock {ICS} security testbed with delay attack case study.
\newblock In \emph{MILCOM 2016-2016 IEEE Military Communications Conference}, pages 283--288. IEEE, 2016.

\bibitem[Li et~al.(2023)Li, Wang, Shen, and Xie]{li2023attack}
Jitao Li, Zhenhua Wang, Yi~Shen, and Lihua Xie.
\newblock Attack detection for cyber-physical systems: A zonotopic approach.
\newblock \emph{IEEE Transactions on Automatic Control}, 2023.

\bibitem[Markovsky et~al.(2022)Markovsky, Huang, and D{\"o}rfler]{markovsky2022data}
Ivan Markovsky, Linbin Huang, and Florian D{\"o}rfler.
\newblock Data-driven control based on the behavioral approach: From theory to applications in power systems.
\newblock \emph{IEEE Control Syst.}, 2022.

\bibitem[Nonhoff and M{\"u}ller(2022)]{nonhoff2022online}
Marko Nonhoff and Matthias~A M{\"u}ller.
\newblock Online convex optimization for data-driven control of dynamical systems.
\newblock \emph{IEEE Open Journal of Control Systems}, 1:\penalty0 180--193, 2022.

\bibitem[Russo(2023)]{russo2023analysis}
Alessio Russo.
\newblock Analysis and detectability of offline data poisoning attacks on linear dynamical systems.
\newblock In \emph{Learning for Dynamics and Control Conference}, pages 1086--1098. PMLR, 2023.

\bibitem[Russo and Proutiere(2021)]{russo2021poisoning}
Alessio Russo and Alexandre Proutiere.
\newblock Poisoning attacks against data-driven control methods.
\newblock In \emph{2021 American Control Conference (ACC)}, pages 3234--3241. IEEE, 2021.

\bibitem[Sakhnini et~al.(2021)Sakhnini, Karimipour, Dehghantanha, and Parizi]{sakhnini2021physical}
Jacob Sakhnini, Hadis Karimipour, Ali Dehghantanha, and Reza~M Parizi.
\newblock Physical layer attack identification and localization in cyber--physical grid: An ensemble deep learning based approach.
\newblock \emph{Physical Communication}, 47:\penalty0 101394, 2021.

\bibitem[Steentjes et~al.(2021)Steentjes, Lazar, and Van~den Hof]{steentjes2021data}
Tom~RV Steentjes, Mircea Lazar, and Paul~MJ Van~den Hof.
\newblock On data-driven control: Informativity of noisy input-output data with cross-covariance bounds.
\newblock \emph{IEEE Control Systems Letters}, 6:\penalty0 2192--2197, 2021.

\bibitem[Teixeira et~al.(2015)Teixeira, Shames, Sandberg, and Johansson]{teixeira2015secure}
Andr{\'e} Teixeira, Iman Shames, Henrik Sandberg, and Karl~Henrik Johansson.
\newblock A secure control framework for resource-limited adversaries.
\newblock \emph{Automatica}, 51:\penalty0 135--148, 2015.

\bibitem[Turan and Ferrari-Trecate(2021)]{turan2021data}
Mustafa~Sahin Turan and Giancarlo Ferrari-Trecate.
\newblock Data-driven unknown-input observers and state estimation.
\newblock \emph{IEEE Control Systems Letters}, 6:\penalty0 1424--1429, 2021.

\bibitem[Van~Waarde et~al.(2020)Van~Waarde, Eising, Trentelman, and Camlibel]{van2020data}
Henk~J. Van~Waarde, Jaap Eising, Harry~L Trentelman, and M~Kanat Camlibel.
\newblock Data informativity: a new perspective on data-driven analysis and control.
\newblock \emph{IEEE Transactions on Automatic Control}, 65\penalty0 (11):\penalty0 4753--4768, 2020.

\bibitem[Van~Waarde et~al.(2023)Van~Waarde, Eising, Camlibel, and Trentelman]{van2023informativity}
Henk~J. Van~Waarde, Jaap Eising, M.~Kanat Camlibel, and Harry~L. Trentelman.
\newblock The informativity approach: To data-driven analysis and control.
\newblock \emph{IEEE Control Systems Magazine}, 43\penalty0 (6):\penalty0 32--66, 2023.

\bibitem[Wang et~al.(2020)Wang, Weng, and Daniel]{wang2020neural}
Yuh-Shyang Wang, Lily Weng, and Luca Daniel.
\newblock Neural network control policy verification with persistent adversarial perturbation.
\newblock In \emph{International Conference on Machine Learning}, pages 10050--10059. PMLR, 2020.

\bibitem[Wigren and Teixeira(2023)]{wigren2023line}
Torbjörn Wigren and André Teixeira.
\newblock On-line identification of delay attacks in networked servo control.
\newblock \emph{IFAC-PapersOnLine}, 56\penalty0 (2):\penalty0 977--983, 2023.
\newblock 22nd IFAC World Congress.

\bibitem[Wu(2022)]{wu2022equivalence}
Liang Wu.
\newblock Equivalence of {SS}-based {MPC} and {ARX}-based {MPC}.
\newblock \emph{arXiv preprint arXiv:2209.00107}, 2022.

\end{thebibliography}
\newpage
\renewcommand{\thesection}{A}
\setcounter{section}{0}  

\section{Appendix}\label{sec:App}
\subsection{Proof of Lemma \ref{lem:data_OL}}\label{app3}
To help us prove Lemma \ref{lem:data_OL}, we next provide two intermediate results.
\begin{lemma}\label{lem:SS_ARX}
Consider a controllable and observable DT LTI SS model defined by the tuple $\Sigma \triangleq (A,B,C,0)$. Let $y[k] \in \mathbb{R}^m, u[k] \in \mathbb{R}^m, x[k] \in \mathbb{R}^n$ represent the output signal, input signal, and state signal. Then there exists an equivalent ARX model denoted by 
    \begin{equation}\label{ARX}
    y[k]+\Upsilon_{n}z[k-1]+\dots+\Upsilon_{1}z[k-n]=\Pi_{n}u[k-1]+\dots + \Pi_{1}u[k-n]. \qquad \square
\end{equation}
\end{lemma}
The proof of Lemma \ref{lem:SS_ARX} is well established (see \citet[Chapter 4]{antsaklis1997linear}, and \citet[Section II.B]{wu2022equivalence}) and thus is omitted. We next recall a result from \citet[Theorem 1]{de2019formulas}.
\begin{lemma}\label{lem:data_de_persis}
Consider a controllable and observable DT LTI SS model defined by the tuple $\Sigma \triangleq (A,B,C,0)$. Let $y[k] \in \mathbb{R}^m, u[k] \in \mathbb{R}^m, x[k] \in \mathbb{R}^n$ represent the output signal, input signal, and state signal. Then, if $\text{rank}\begin{bmatrix}
    U_{0,1,T}\\X_{1,T}
\end{bmatrix}=n+m$, then the state space model has an equivalent representation 
\begin{equation}
    x[k+1] = X_{1,T}\begin{bmatrix}
    U_{0,1,T}\\
        X_{0,T}
    \end{bmatrix}\ssymbol{2} \begin{bmatrix}
    u[k]\\
        x[k]
    \end{bmatrix}. \quad \square
\end{equation}
\end{lemma}
Using the above two results, we now begin to prove Lemma \ref{lem:data_OL}.

\textit{Proof of Lemma \ref{lem:data_OL}:} Consider the SS model in \eqref{eq:SS_con_form}. Using the results of Lemma \ref{lem:SS_ARX}, we know that for the SS model \eqref{eq:SS_con_form}, there exists an equivalent ARX model as follows
 \begin{equation}\label{ARX}
    z_j[k]+\Upsilon_{j,n}z_j[k-1]+\dots+\Upsilon_{j,1}z_j[k-n]=\Pi_{{j,n}}u[k-1]+\dots + \Pi_{j,1}u[k-n],\; \forall j \in \Omega
\end{equation}
where $\Omega := \{1, 2,\dots, {N \choose N-M}\}$Using the definition of states $\mathcal{X}_j$ in \eqref{data_matrices0}, we formulate an equivalent state space formulation as
\begin{equation}\label{eq:CL}
\mathcal{X}_j[k+1]=\bar{A}_j\mathcal{X}_j[k] +\bar{B}u[k],
\end{equation}
where $\setlength{\dashlinegap}{0.8pt}\left[\begin{array}{c:c}
  \bar{A}_j   & \bar{B}
\end{array}\right] \triangleq $
\begin{equation}\label{eq:A_B}
\setlength{\dashlinegap}{0.8pt}
\left[\begin{array}{cccccccccc:c}
0 & I & 0 & \dots & 0 & 0 & 0 & 0 & \dots & 0 & 0\\
0 & 0 & I & \dots & 0 & 0 & 0 & 0 & \dots & 0 & 0\\
\vdots & \vdots & \vdots & \ddots & \vdots & \vdots & \vdots & \vdots & \ddots & \vdots  & \vdots\\
0 & 0 & 0 & \dots & I & 0 & 0 & 0 & \dots & 0 & 0\\
-\Upsilon_{j,n}& -\Upsilon_{j,n-1} & -\Upsilon_{j,n-2} & \dots & -\Upsilon_{j,1} & \Pi_{j,n} & \Pi_{j,n-1} & \Pi_{j,n-2} & \dots & \Pi_{j,1}  & 0\\
\hdashline
0 & 0 & 0 & \dots & 0 & 0 & I  & 0 & \dots & 0 & 0\\
0 & 0 & 0 & \dots & 0 & 0 & 0  & I & \dots & 0 & 0\\
\vdots & \vdots & \vdots & \ddots & \vdots & \vdots & \vdots & \vdots & \ddots & \vdots & \vdots\\
0 & 0 & 0 & \dots & 0 & 0 & 0  & 0 & \dots & I & 0\\
0 & 0 & 0 & \dots & 0 & 0 & 0  & 0 & \dots & 0 & I
\end{array}\right]
\end{equation}
Note here that, the input-output data that was collected in $\mathcal{I}_T$ obeys the dynamics in \eqref{eq:CL}. In other words, we have now constructed a state space representation with access to input and state data. Then we conclude the proof using Lemma \ref{lem:data_de_persis}. $\hfill \blacksquare$
\subsection{Proof of Lemma \ref{lem:rank_obsv}}\label{app2}
Let $Q:=N-M$. For any given $j \in \left\{1,2,\dots,{N \choose Q}\right\}$, it holds from \eqref{data_matrices0} that 
\begin{equation}\label{e1}
    \underbrace{\begin{bmatrix}
        U_{n,T}\\
        \hline
        \hat{X}_{j,n,T}
    \end{bmatrix}}_{M_0}=\begin{bmatrix}
        U_{n,T}\\
        \hline 
        Z_{j,0,n,T}\\
        U_{0,n,T}
    \end{bmatrix}, \quad 
    \text{where}\quad 
    Z_{j,0,n,T} = \underbrace{\begin{bmatrix}
         \mathcal{O}_{j,n} & \mathcal{T}_{j,n}
    \end{bmatrix}}_{M_1} \underbrace{\begin{bmatrix}
        X_{0,T}\\
        U_{0,n,T}
    \end{bmatrix}}_{M_2}
\end{equation}
where the matrix $X_{0,T}$ is defined similarly to \citet[(4)]{de2019formulas}, and $
\left[\begin{array}{c|c}
\mathcal{O}_{j,n} & \mathcal{T}_{j,n}
\end{array}\right]=$
\begin{equation}\label{e0}
\left[\begin{array}{c|cccc}
        C_{{\mathcal{J}}_{j}} & 0 & 0 & \dots & 0\\
        C_{{\mathcal{J}}_{j}}A & C_{{\mathcal{J}}_{j}}B & 0& \dots & 0\\
        C_{{\mathcal{J}}_{j}}A^2 & C_{{\mathcal{J}}_{j}}AB & C_{{\mathcal{J}}_{j}}B& \dots & 0\\
        \vdots & \vdots & \vdots & \ddots & \vdots\\
        C_{{\mathcal{J}}_{j}}A^{n-1} & C_{{\mathcal{J}}_{j}}A^{n-2}B & C_{{\mathcal{J}}_{j}}A^{n-3}B & \dots & 0
\end{array}\right]
\end{equation}
The matrix $M_0$ is of dimension $M_0 \in \mathbb{R}^{s \times T}$ where $s \triangleq m(n+1)+Qn$. We aim to show that condition \eqref{eq:PE}, or equivalently the condition 
\begin{equation}\label{rank}
\text{rank}(M_0)=s,
\end{equation}
does not hold when the tuple $(A,C_{{\mathcal{J}}_{j}})$ is not observable. From the theorem statement, since we know that $T > m(n+1)+Qn$, we observe that $M_0$ is a fat matrix (more columns than rows). Thus, the condition \eqref{rank} holds only if $M_0$ has full row rank $s$. From \eqref{e1}, one of the necessary conditions for \eqref{rank} to hold is
\begin{equation}\label{e3}
    \text{rank}(Z_{j,0,n,T})= nQ. 
\end{equation}
Using \eqref{e1}, \eqref{e3} is rewritten as 
\begin{equation}\label{f1}
 \text{rank}(Z_{j,0,n,T})= \min \{\text{rank}(M_1),  \text{rank}(M_2)\} =nQ
\end{equation}
From \citet[Lemma 1]{de2019formulas} we know that for any controllable system where the input is persistently exciting of order $(m+Q)n+1$, it holds that $\text{rank}(M_2) = n(m+1)$, i.e., $M_2$ is of full rank. Then \eqref{f1} can be re-writtten as
\begin{equation}\label{e5}
\min \{\text{rank}(M_1),n(m+1)\} = nQ
\end{equation}
where $M_1 \in \mathbb{R}^{Qn \times n(m+1)}$. We next show that condition \eqref{e5} (which is a necessary condition for \eqref{e3} to hold) cannot hold when the tuple $(A,C_{{\mathcal{J}}_{j}})$ is not observable. 

First, consider the case that $Q > m+1$. Then $\text{rank}(M_0) \neq nQ$ since the rank cannot be larger than the smallest dimension of the matrix. Second, consider the case that $Q = m+1$. Then $\text{rank}(M_0) = nQ$, only if the matrix $M_0$ has full rank. However, we observe that there is a zero column in $M_0$ which reduces the rank. Finally, we consider the case that $Q< m+1$. Then, 
if the tuple $(A,C_{{\mathcal{J}}_{j}})$ is not observable, $\text{row\;rank}(\mathcal{O}_{j,n}) \leq n-1$. Let this be observation \textbf{(O1)}.

Consider the matrix $\mathcal{T}_{j,n} \in \mathbb{R}^{nQ \times nm}$. Since one of the rows of $\mathcal{T}_{j,n}$ is zero, the maximum row rank that the matrix $\mathcal{T}_{j,n}$ can attain is $nQ-Q$. Let this be observation \textbf{(O2)}. 

Combining \textbf{(O1)} and \textbf{(O2)}, the maximum row rank of $M_1$ is the sum of the row ranks of $\mathcal{O}_{j,n}$ and $\mathcal{T}_{j,n}$. That is, $\max \text{rank}(M_1) \leq n-1+nQ-Q$. Plugging this inequalities in the L.H.S of \eqref{e5}, we get
\begin{equation}\label{e6}
    \text{rank}(Z_{j,0,n,T}) = \min\{nQ+n-Q-1, n(m+1)\},
\end{equation}
From the Lemma statement, $n < Q+1 \implies n-Q-1 <0$, then 
\begin{equation}\label{e7}
 nQ > nQ+n-1-Q.    
\end{equation}
Substituting \eqref{e7} into \eqref{e6}, we get $\text{rank}(Z_{j,0,n,T}) < nQ$
which concludes the proof. $\hfill \blacksquare$
\subsection{Proof of Proposition \ref{prop1}}\label{app5}
Let $Q:= N-M$ and consider the data-based representation of the plant in \eqref{data:OL} which is attack free (Line $1$ of Algorithm \ref{alg1}). Since the sensors are attack free for the first $n$ time instants, we can construct an attack free matrix $\underline{\hat{X}}_{j,k,1}$ (Line $2$ of Algorithm \ref{alg1}). Since the model $\Lambda_j$, the initial condition $\underline{\hat{X}}_{j,k,1}$, and the input applied $u[k]$ are attack-free, the predicted output from the data-based representation \eqref{data:OL} will be attack-free. However, if the output at time instant $k+1$ is under attack, the norm in \eqref{eq:attack_cond} will be non-zero due to the difference in predicted outputs and received outputs. Since \eqref{eq:attack_cond} searches for the least value of the difference between predicted and received output values (Line $3$ of Algorithm \ref{alg1}), the minimum will not occur for any sensors  $j \in \{1, 2, \dots, {N\choose Q}\}$ under attack. If no sensors are under attack, then we update our initial condition using the received output and go back to the prediction step (Line $3$ of Algorithm \ref{alg1}). Since the received output was attack-free (since all the sensors were deemed to be attack-free), the updated initial condition will also be attack free. This concludes the proof. $\hfill \blacksquare$
\subsection{Proof of Lemma \ref{lem:replay}}\label{App6}
Consider the plant under replay attacks \eqref{replay}. Now, let us consider the rank condition 
\begin{equation}\label{t1}
    \text{rank}\begin{bmatrix}
        \underline{U}_{n,T_1},\\ \underline{\hat{X}}_{j,n,T_1},
    \end{bmatrix}=m(n+1)+Qn,
\end{equation}
As shown in the proof of Lemma \ref{lem:rank_obsv}, a necessary condition  for \eqref{t1} to hold is
\begin{equation}\label{t2}
    \text{rank}(\underline{Z}_{j,0,n,T_1})= nQ, 
\end{equation}
when $k=0$.The matrix $\underline{Z}_{j,0,n,T_1}$ is a fat matrix and satisfies \eqref{t2} only if it has full row rank. To inspect the row rank, let us rewrite $\underline{Z}_{j,0,n,T_1}$ as 
\begin{equation}
\begin{bmatrix}
        z_j[0]\vert_{1} & z_j[1]\vert_1 & \dots & z_j[T_1-1]\vert_{1}\\
        z_j[1]\vert_{1} & z_j[2]\vert_{1} & \dots & z_j[T_1]\vert_{1}\\   
        \vdots & \vdots & \ddots & \vdots\\
        z_j[n-1]\vert_{1} & z_j[n]\vert_{1} & \dots & z_j[n-1+T_1]\vert_{1}
    \end{bmatrix}
\end{equation}
where $\vert_{1}$ denotes that under replay attack, at least one of the rows of $z_j[\cdot]$ is a constant $c_j$. Then at least $n$ rows of the matrix are constants which makes the matrix row rank deficient. This concludes the proof. $\hfill \blacksquare$

\subsection{Proof of Proposition \ref{prop:replay}}\label{App7}
Let us apply persistently exciting inputs of order $\tilde{n}$ to the system and construct the corresponding output matrices  (Line $1$ of Algorithm \ref{alg:replay}). We know from Lemma \ref{lem:replay} that the rank condition \eqref{rank:replay} cannot be satisfied under replay attacks. Thus, if any sensor pair $j \in \{1, 2,\dots, {N \choose N-M}\}$ is under attack, the rank condition \eqref{rank:replay} fails. 

We know from \citet[Lemma 1]{de2019formulas} that for any controllable system, the rank condition \eqref{rank:replay} is satisfied if the input is persistently exciting of order $\tilde{n}$. We know from Assumption \ref{assum:q_sen_obsv} that there exists a sensor pair $j \in \{1, 2,\dots, {N \choose N-M}\}$ which is replay attack free and thus would satisfy the rank condition \eqref{rank:replay}. Since line $3$ of Algorithm \ref{alg:replay} determines the sensor pair which maximizes the rank, the sensors which are attack-free would achieve the maximum rank which also concludes the proof. $\hfill \blacksquare$
\subsection{Proof of Proposition \ref{prop:delay}}\label{App8}
Let us apply a non-zero input at any given time instant $k$ (Line $1$ of Algorithm \ref{alg:delay}). Then, for any given attack-free sensor $j \in \{1, 2, \dots, N\}$, the first non-zero output occurs at time instant $r_j$. If the sensor $j$ is under delay attack, the time instant at which the first non-zero output occurs will be larger than $r_j$. And equation \ref{eq:att_delay} determines the sensor which has the minimum difference between the first non-zero output and the relative degree $r_j$ (Line $2$ of Algorithm \ref{alg:delay}). From the statement of the Proposition, we know that one of the sensors is attack free. Thus, the minimum in \eqref{eq:att_delay} will be zero and will correspond to the attack-free sensor. This concludes the proof. $\hfill \blacksquare$
\end{document}